\documentclass[pra,twocolumn,showpacs,preprintnumbers]{revtex4}
\usepackage{amsmath}
\usepackage{graphicx}
\usepackage{mathrsfs}
\usepackage{amssymb}
\usepackage{amsfonts}
\usepackage{amsmath}
\usepackage[colorlinks,citecolor=blue,linkcolor=red,hyperindex,CJKbookmarks,dvipdfm]{hyperref}

\begin{document}

\title{Multi-ion Mach-Zehnder interferometer with artificial nonlinear
interactions}

\author{Y. M. Hu$^{1,2,4}$, W. L. Yang$^{1}$, X. Xiao$^{3}$}

\author{M. Feng$^{1}$}
\altaffiliation{Corresponding author. Email: mangfeng@wipm.ac.cn}

\author{C. Lee$^{4}$}
\altaffiliation{Corresponding author. Email: chleecn@gmail.com}

\affiliation{$^{1}$State Key Laboratory of Magnetic Resonance and Atomic and Molecular Physics, Wuhan Institute of Physics and Mathematics, Chinese Academy of Sciences, and Wuhan National Laboratory for Optoelectronics, Wuhan 430071, China}

\affiliation{$^{2}$Graduate School of the Chinese Academy of Sciences, Beijing 100049, China}

\affiliation{$^{3}$College of Physics and Information Science, Hunan Normal University, Changsha 410081, China}

\affiliation{$^{4}$State Key Laboratory of Optoelectronic Materials and Technologies, School of Physics and Engineering, Sun Yat-Sen University, Guangzhou 510275, China}

\begin{abstract}
We show how to implement a Mach-Zehnder interferometry based upon a string of trapped ions with artificial nonlinear interactions. By adiabatically sweeping down/up the coupling strength between two involved internal states of the ions, we could achieve the beam splitting/recombination. Based on current techniques for manipulating trapped ions, we discuss the experimental feasibility of our scheme and analyze some undesired uncertainty under realistic experimental environment.
\end{abstract}

\maketitle

\section{introduction}

Due to long coherence time of some specific hyperfine states, high
controllability of operations and high efficiency of state detection, the
trapped atomic ion system has been considered as a promising candidate for
quantum information processing. In particular, since they could be nearly
perfectly prepared in some entangled states, the trapped ions have attracted
considerable attention for high-precision quantum metrology~\cite{ion1,
ion2, Wineland, Wineland1, Leibfried}.

It is generally believed that the measurement precision can be enhanced from
the standard quantum limit (SQL) or shot noise limit to the Heisenberg limit
by utilizing multipartite entanglement~\cite{Giovannetti}. For example, it
has demonstrated that the measurement precision may reach the Heisenberg
limit by using a NOON state $\left( \left\vert N\right\rangle _{a}\left\vert
0\right\rangle _{b}+\left\vert 0\right\rangle _{a}\left\vert N\right\rangle
_{b}\right) /\sqrt{2}$, which is a superposition of N particles in mode $a$
with zero particle in mode $b$ and vice versa. However, most of the relevant
proposals on high-precision interferometry of trapped ions are subject to
limited numbers of ions or to the requirement for individual addressing of
ions, which restricts the scalability of those interferometry schemes.

In this article, we consider a Mach-Zehnder (MZ) interferometry using large
numbers of entangled trapped ions. The MZ interferometry consists of a beam
splitter for splitting the incoming state and another beam splitter for
state recombination. It has been shown that the adiabatic MZ interferometry
is an optimal candidate for high-precision measurement~\cite{Lee2006,Lee2009}%
. To achieve such an interferometry with trapped ions, we have to realize a
nonlinear giant-spin Hamiltonian with anisotropic interaction. The key point
is how to generate the nonlinear interactions and then achieve the beam
splitting and recombination. By applying a pair of lasers to manipulate
independently their center-of-mass (COM) modes (either transverse or
longitudinal vibrational modes), the nonlinear giant-spin Hamiltonian with
anisotropic interaction could be simulated bases upon an array of trapped
ions in a linear trap. By adjusting the coupling strength (i.e. the Rabi
frequency) between the two involved hyperfine states, the MZ interferometry
could be carried out adiabatically. With current techniques for manipulating
trapped ions, we discuss the experimental feasibility of our scheme and the
difficulty due to decoherence in realistic experimental environment.

The paper is structured as follows. In the next section, we describe the
physical system and derive its effective Hamiltonian. In Sec. III, we show
how to realize the MZ interferometer by using quantum adiabatic processes
through dynamical bifurcations. Then, we discuss the experimental
feasibility in Sec. IV and analyze the undesired uncertainty due to
decoherence in Sec. V. In the last section, we briefly summarize our results.

\begin{figure*}[tbp]
\includegraphics[width=5.3in,height=1.5in]{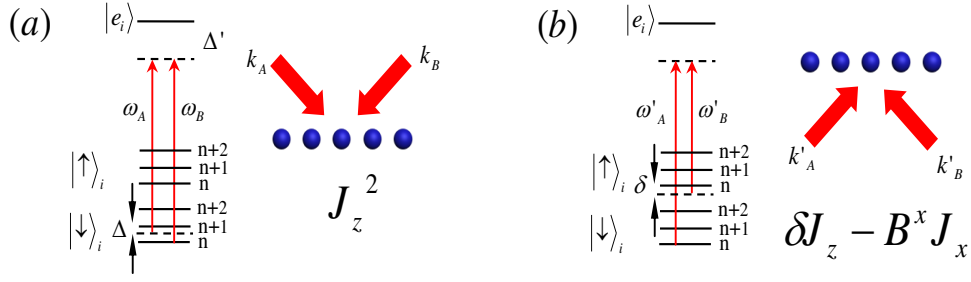}
\caption{(color online) Laser coupling scheme for simulating the Hamiltonian
(2). (a) Physical realization of the nonlinear term of J$_{z}^{2}$: a Stark
shift regarding the level $\left\vert \downarrow \right\rangle _{i}$ is
created if two laser beams uniformly radiate the ions for the excitation
from $\left\vert \downarrow \right\rangle _{i}$ to $\left\vert
e\right\rangle _{i}$ with a large detuning $\Delta ^{\prime }$. (b) Physical
realization of the terms of J$_{x}$ and J$_{z}$: two additional laser beams
are used to induce the carrier transitions corresponding to the term of J$_{x}$;
and therefore the detuning to the resonant transition $\delta=\protect\omega _{0}-\left(\protect\omega _{A}^{\prime }-\protect\omega _{B}^{\prime }\right)$ will generate the
longitudinal-field term $\protect\delta J_{z}$. See more details in Appendix
A.}
\label{fig:wide}
\end{figure*}

\section{Physical system and its effective Hamiltonian}

We consider N identical ions confined in a linear trap, as shown in Fig. 1,
where the three ionic states under our consideration are denoted by two
hyperfine ground spin states ($\left\vert \downarrow \right\rangle $, $%
\left\vert \uparrow \right\rangle $) and an excited state $\left\vert
e\right\rangle $, respectively. In our scheme, all the ions are irradiated
simultaneously by two traveling-wave laser beams with different frequencies,
which coherently couple the hyperfine ground states through the optically
stimulated Raman transitions by adiabatically eliminating the excited state $%
\left\vert e\right\rangle $. Here we assume the detuning of the optical
fields from the electronic resonance to be much larger than the excited
state linewidth and the corresponding coupling strength. As indicated in
Fig. 1(a), under an usual rotating wave approximation to the laser
frequencies, the interaction Hamiltonian between the fields and the ions is
written as ($\hbar =1$)~\cite{job}
\begin{equation}
H_{1}=\nu a^{\dag }a-\frac{1}{\sqrt{N}}\sum_{i=1}^{N}\Omega _{z}\eta
_{z}[(a^{\dag }+a)\sigma _{z}^{i}e^{-i(\nu -\Delta )t}+h.c.],
\end{equation}%
where $\sigma _{z}^{i}=\left\vert \downarrow \right\rangle _{i}\left\langle
\downarrow \right\vert -\left\vert \uparrow \right\rangle _{i}\left\langle
\uparrow \right\vert $ is the population inversion operator for the $i$-th
ion, $a^{\dag }$ $(a)$ is the creation (annihilation) operator of the
center-of-mass (COM) mode and $\Omega _{z}$ is the uniform effective Rabi
frequency. $\nu $ is the vibrational frequency of the COM mode and $\Delta $
is the detuning from the vibrational frequency. $\eta _{z}$ denotes the
Lamb-Dicke parameter $k_{z}\sqrt{\hbar /2M\nu }$ with the ion mass $M$ and
the laser wavevector difference $k_{z}=k_{B,z}-k_{A,z}$.

By tuning frequency differences and radiation directions of the pair of
lasers along with independent manipulation of transverse or longitudinal COM
modes, we may obtain following effective Hamiltonian under some
approximations and canonical transformations,
\begin{equation}
H_{eff}=\delta J_{z}-B^{x}J_{x}-\lambda J_{z}^{2},
\end{equation}
where the collective angular momentum for all the ions are defined as $%
J_{k}=\sum\nolimits_{i=1}^{N}\sigma _{k}^{i}/2$ with $k=(x, y, z)$, the
longitudinal field $\delta$ is the internal energy gap between the two spin
states ($\left\vert \downarrow \right\rangle$, $\left\vert \uparrow
\right\rangle$), and the transverse field $B^{x}$ corresponds to the
effective Rabi frequency, and the effective spin-spin nonlinear interaction
is given as $\lambda =8\Omega _{z}^{2}\eta _{z}^{2}/(N\Delta)$. More details
of our derivation are given in Appendix A. These parameters $\delta$, $B^{x}$%
, and $\lambda $ could be controlled by Rabi frequencies, wavevector
differences of the lasers and the detunings.

\section{MZ interferometry via adiabatic operations}

The MZ interferometry consists of a beam splitter for splitting input states
and another beam splitter for recombining output states. Based upon the
trapped-ion system (2), the two beam splitters for a MZ interferometry could
be achieved by adiabatic quantum evolution through dynamical bifurcations.

\begin{figure}[tbh]
\begin{center}
\includegraphics[width=1.0 \columnwidth]{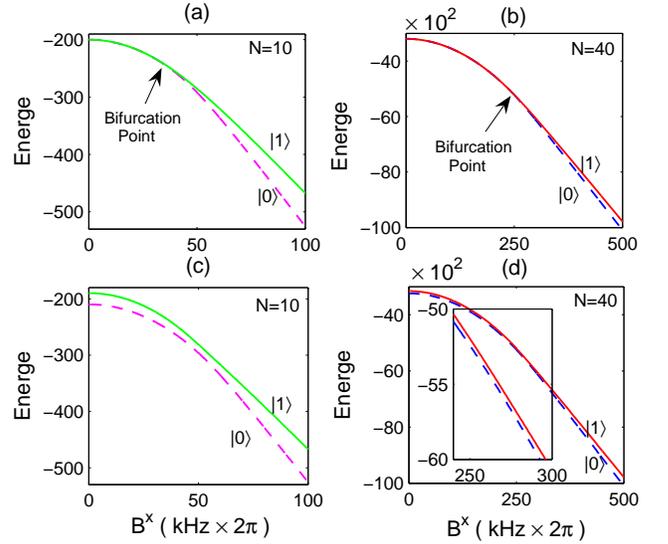}
\end{center}
\caption{(Color online) The energy spectra of the ground state $\left\vert
0\right\rangle $ (dashed line) and the first excited state $\left\vert
1\right\rangle $ (solid line) with respect to $B^{x}$ for different values
of $\protect\delta $ and $N$. If $\protect\delta =0$, there is a bifurcation
from degeneracy to non-degeneracy of the two lowest states $\left\vert
0\right\rangle $ and $\left\vert 1\right\rangle $ when $B^{x}$ increases,
where (a) N=10 and (b) N=40. The beam splitting can be achieved by adiabatic
passage through such a bifurcation. If $\protect\delta \neq 0$, the
degeneracy breaks down even in the weak limit of $B^{x}$, where $\protect%
\delta =\protect\lambda /4$, with N=10 (c) and N=40 (d). The other
parameters are chosen as $\protect\eta _{z}=0.1$, $\Delta =0.01$ MHz, $%
\Omega _{z}=2\protect\pi \times $100 kHz $(N=10)$ and $2\protect\pi \times $%
200 kHz$(N=40)$.}
\label{Fig.2}
\end{figure}

In the collective spin representation, the trapped-ion system is regarded as
an ensemble of $N$ spin-$1/2$ particles. Let $\left\vert jm\right\rangle$
stand for the joint eigenstate of the SU(2) Casimir operators $J^{2}$ and $%
J_{z}$, which satisfy the relations
\begin{equation}
\begin{aligned} J^{2}\left\vert jm\right\rangle =j(j+1)\left\vert
jm\right\rangle, \notag \\ J_{z}\left\vert jm\right\rangle =m\left\vert
jm\right\rangle, \notag \end{aligned}
\end{equation}
with $j=\frac{N}{2}$ and $m=\frac{N}{2},\frac{N}{2}-1,...,-\frac{N}{2}$~\cite%
{Dicke}. $\left\vert \frac{N}{2},-\frac{N}{2}\right\rangle$ ($\left\vert
\frac{N}{2},\frac{N}{2}\right\rangle$) means all ions in the hyperfine state
$\left\vert \downarrow \right\rangle$ ($\left\vert \uparrow \right\rangle$).

We first consider the case $\delta =0$. In the strong coupling limit $%
B^{x}\gg \left\vert \lambda \right\vert $, the ground state $\left\vert
0\right\rangle $ and the first excited state $\left\vert 1\right\rangle $
for the Hamiltonian $H_{eff}$ are non-degenerated, where $\left\vert
0\right\rangle =\exp (i\frac{\pi }{2}J_{y})\left\vert j=\frac{N}{2},m=\frac{N%
}{2}\right\rangle $ is a spin coherent state, and $\left\vert 1\right\rangle
$ is a superposition of different states $\left\vert jm\right\rangle $. When
we adiabatically tune $B^{x}$ from the strong coupling limit ($B^{x}\gg
\left\vert \lambda \right\vert $) to the weak coupling limit ($B^{x}\ll
\left\vert \lambda \right\vert $), the ground state $\left\vert
0\right\rangle $ and the first excited state $\left\vert 1\right\rangle $
evolve into the states $\left\vert \frac{N}{2},-\frac{N}{2}\right\rangle $
and $\left\vert \frac{N}{2},\frac{N}{2}\right\rangle $, respectively. If $%
B^{x}=0$, the two lowest states turn to be degenerated. So in such an
adiabatic passage, there exists a transition from non-degeneracy to
degeneracy between $\left\vert 0\right\rangle $ and $\left\vert
1\right\rangle $, which can be regarded as a bifurcation~\cite%
{Lee2004,ex-bifurcation}, as shown in (a) and (b) of Fig. 2. Similar to a
single-particle MZ interferometry, where the ground state $\left\vert
0\right\rangle $ and the first exited state $\left\vert 1\right\rangle $ are
utilized as two paths, the two states of maximum spin $\left\vert \frac{N}{2}%
,-\frac{N}{2}\right\rangle $ and $\left\vert \frac{N}{2},\frac{N}{2}%
\right\rangle $ can also be used as two paths for an N-particle MZ
interferometry. Therefore, the achievement of the maximally path-entangled
state $\left\vert \Psi _{p}\right\rangle =\frac{1}{\sqrt{2}}\left(
\left\vert \frac{N}{2},\frac{N}{2}\right\rangle +\left\vert \frac{N}{2},-%
\frac{N}{2}\right\rangle \right) $ implies accomplishment of the beam
splitting. Then we turn off the lasers for a duration $T$ so that the
maximally path-entangled state $\left\vert \Psi _{p}\right\rangle $ evolves
into
\begin{equation}
\left\vert \Psi _{p}^{^{\prime }}\right\rangle =\frac{1}{\sqrt{2}}\left(
e^{-i\frac{N}{2}\omega _{0}T}\left\vert \frac{N}{2},\frac{N}{2}\right\rangle
+e^{i\frac{N}{2}\omega _{0}T}\left\vert \frac{N}{2},-\frac{N}{2}%
\right\rangle \right) ,
\end{equation}%
where $\omega _{0}$ is the frequency difference between the two hyperfine
spin states $\left\vert \downarrow \right\rangle $ and $\left\vert \uparrow
\right\rangle $.

There are two different methods for extracting the relative phase
information between the two paths of the N-particle MZ interferometry. In
the case of few trapped ions, this could be carried out using controlled-NOT
gates $CNOT_{rest}^{1}$ with the first ion being the control qubit and the
rest ions being the target ones, followed by a Hadamard operation ($H_{1}$)
on the first ion~\cite{Wineland}. The relative phase information can be
obtained by measuring the population of the first ion in the state $%
\left\vert \downarrow \right\rangle$ or $\left\vert \uparrow \right\rangle$.
Taking $N=5$ as an example, this individually addressing method can be
summarized briefly as follows,
\begin{eqnarray}
&&\left\vert \Psi _{p}^{^{\prime }}\right\rangle \frac{}{}^{\underrightarrow{%
N=5~}}\frac{1}{\sqrt{2}}\left(e^{-i\frac{5}{2}\omega _{0}T}\left\vert
\uparrow \uparrow \uparrow \uparrow \uparrow \right\rangle +e^{i\frac{5}{2}%
\omega _{0}T}\left\vert \downarrow \downarrow \downarrow \downarrow
\downarrow \right\rangle \right)  \notag \\
&&\frac{}{}^{\underrightarrow{CNOT_{rest}^{1}~}}\frac{1}{\sqrt{2}}\left(e^{-i%
\frac{5}{2}\omega_{0}T} \left\vert \uparrow \downarrow \downarrow \downarrow
\downarrow\right\rangle +e^{i\frac{5}{2}\omega _{0}T}\left\vert \downarrow
\downarrow\downarrow \downarrow \downarrow \right\rangle \right)  \notag \\
&&\frac{}{}^{\underrightarrow{H_{1}~}}\cos \left(\frac{5}{2}\omega
_{0}T\right)\left\vert \downarrow \downarrow \downarrow \downarrow
\downarrow \right\rangle +i\sin \left(\frac{5}{2}\omega_{0}T\right)\left%
\vert \uparrow \downarrow \downarrow \downarrow \downarrow \right\rangle.
\end{eqnarray}

However, for systems of large numbers of ions, individual addressing becomes
challenging and it becomes very difficult to extract the relative phase with
the controlled-NOT gate $CNOT_{rest}^{1}$. If we apply the inverse process
of the first beam splitter for recombing the two paths, the relatives phase
information indeed could be transferred into the population information of
the two lowest states for the limit of $B^{x}\gg \lambda $. However, it is
not easy to directly distinguish these two lowest states. Fortunately, the
two lowest states for symmetric ($\delta =0$) and asymmetric ($\delta \neq 0$%
) systems of $B^{x}\ll \lambda $ are almost identical and the two lowest
states ($\left\vert \frac{N}{2},-\frac{N}{2}\right\rangle $, $\left\vert
\frac{N}{2},\frac{N}{2}\right\rangle $) become non-degenerate for an
asymmetric system ($0<\delta <\lambda /4$), see Fig. 2 (c) and (d).
Therefore, to extract the relative phase, after applying the inverse process
of the beam splitting process, we suddenly apply a proper nonzero $\delta $ (%
$0<\delta <\lambda /4$) and then adiabatically decrease $B^{x}$ from $%
B^{x}\gg \lambda $ to $B^{x}=0$. This recombination procedure for extracting
the relative phase with two adiabatic processes could be summarized as,
\begin{widetext}
\begin{eqnarray}
&&\left\vert \Psi _{p}^{^{\prime}}\right\rangle \frac{}{}^{\underrightarrow{\text{adiabatic process of } \delta =0, B^{x}=0 \rightarrow B^{x}\gg \lambda~~}} \frac{\cos (N\omega_{0}T/2)\left\vert 0\right\rangle-i\sin (N\omega _{0}T/2)\left\vert1\right\rangle}{\sqrt{2}}  \notag \\
&&^{\underrightarrow{\text{adiabatic process of } \delta \ne 0, B^{x}\gg \lambda \rightarrow B^{x}= 0~~}}\text{\ }\frac{\cos (N\omega _{0}T/2)\left\vert \frac{N}{2},\frac{N}{2}\right\rangle -i\sin (N\omega_{0}T/2)\left\vert \frac{N}{2},-\frac{N}{2}\right\rangle }{\sqrt{2}}.
\end{eqnarray}
\end{widetext}Obviously, by measuring the population with all the ions in
the state $\left\vert \downarrow \right\rangle $ or $\left\vert \uparrow
\right\rangle $, the relative phase between the two paths could be obtained.

\section{Experimental feasibility}

Our scheme is feasible with current ion trap technology. If we employ a line
of trapped $^{40}$Ca$^{+}$, the states $\left\vert \downarrow \right\rangle $%
, $\left\vert \uparrow \right\rangle $ and $\left\vert e\right\rangle $ are
denoted by S$_{1/2}(m_{j}=-1/2)$, S$_{1/2}(m_{j}=1/2)$, and P$%
_{1/2}(m_{j}=-1/2)$~\cite{steane}. To adiabatically eliminate the ionic
excited state $\left\vert e\right\rangle $, the detuning of the optical
fields from electronic resonance should be much larger than the excited
state linewidth and the corresponding coupling strength, e.g. $\Delta
^{\prime }=2\pi \times 20$ GHz. On the other hand, generation of the
nonlinear term $\lambda J_{z}^{2}$ in the Hamiltonian (2) requires that the
detuning $\Delta $ should be much smaller than the frequency $\nu $ of the
longitudinal COM mode. As a result, in our case we may take $\nu =1\;$MHz
and $\Delta =0.01\nu $. Besides, the coherence time of the motional states
has reached 100 ms \cite{cohererent}. To change the effective coupling $%
B^{x} $ adiabatically, we may set $B^{x}(t)=\alpha -\beta t$, where $\alpha $
and $\beta $ are nonnegative parameters, for example, $\alpha =500$ kHz and $%
\beta =50$ kHz/ms with $0\leq t\leq 10$ $ms$ in the case of $N=40$. Consider
the system in a free evolution by $T=5$ ms. After a duration for about 35 ms
including three adiabatic processes plus the free evolution, we could obtain
the relative phase by measuring the population of all the ions
in state $\left\vert \frac{N}{2},-\frac{N}{2}\right\rangle$ or
$\left\vert \frac{N}{2},+\frac{N}{2} \right\rangle$.

Higher measurement precision for the trapped-ion MZ interferometer requires
more ions involved. With more ions involved, however, besides more
attentions should be paid to larger coupling strength, we also need to
consider the detrimental influence from decoherence. Fortunately, currently
available technologies enable the effective coupling strength to be on the
order of MHz, which meets the requirement of our proposal. Therefore, we
focus below on the influence from decoherence.

\section{Estimation of frequency uncertainty under decoherence}

There are several sources of decoherence. Here, as an example, we only
consider the depasing decoherence influence on the estimation of the ionic
Zeeman splitting $\omega _{0}$. The maximally entangled state of $N$ trapped
ions undertaking MZ interferometry decreases the measurement uncertainty of $%
\omega _{0}$ by $\sqrt{N}$ times compared to the trapped ions in product
states~\cite{Wineland,Leibfried}. If decoherence due to environment is
involved, the measurement precision would be seriously affected~\cite%
{estimation1,estimation2} because entangled states are more sensitive to
decoherence than product states. A recent experiment~\cite{correlated} has
reported that the NOON state with 8 to 14 trapped ions subjects to
correlated dephasing noise mostly because of magnetic field fluctuation and
shows super-decoherence, where the coherence decays quadratically with the
number of qubits. This correlated dephasing could be modeled by the Lindblad
equation in the free evolution,
\begin{equation}
\dot{\rho}=-i\omega _{0}[J_{z},\rho ]+\gamma (2J_{z}\rho J_{z}-J_{z}^{2}\rho
-\rho J_{z}^{2})
\end{equation}%
where $\gamma $ is the correlated dephasing rate, which is $N^{2}\gamma _{0}$
with $\gamma _{0}$ the dephasing rate of a single qubit. Here, we consider $%
\gamma _{0}=$ $\gamma _{0}^{\prime }\Omega _{z}^{2}/\Delta ^{\prime 2}$ to
be the effective two-level dephasing rate~\cite{effective} due to the
two-photon Raman process, where $\gamma _{0}^{\prime }$ is the spontaneous
decay rate of the excited state $\left\vert e\right\rangle $ for a single
qubit. The uncertainty $\Delta \omega /\omega _{0}$ shown in Fig. 3 is
calculated in the equivalent spin representation, where the details are
given in Appendix B.

Repeating the implementation of our proposal by one hundred times, we find
that the correlated dephasing seriously spoils the measurement precision of
the entangled state, yielding the precision even lower than the SQL. By
setting $\gamma _{0}^{\prime }\approx 2\pi\times 20$ MHz, $\Omega _{z}=2\pi
\times 200$ kHz, and $\Delta^{\prime }=2\pi\times 20$ GHz, and considering
free evolution by $T=5$ ms in each implementation, we have $\gamma T=0.1$
and the uncertainty of $\omega_{0}$ in the case of three-ion entanglement is
larger than the SQL, see Fig. 3. In order to enhance the precision, we have
to reduce $\gamma$. A possible way is to enlarge $\Delta ^{\prime }$ by ten
times, which leads to $\gamma T$ to be about 0.001, and thereby the
measurement precision would approach the Heisenberg limit. Further
improvement could be made by using refocusing pulses during the adiabatic
operations to suppress the decay from $|e\rangle$.

\begin{figure}[tbh]
\begin{center}
\includegraphics[width=0.8 \columnwidth]{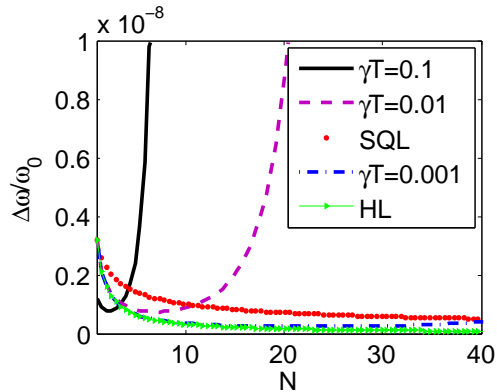}
\end{center}
\caption{(color online) The frequency uncertainty $\Delta \protect\omega /%
\protect\omega _{0}$ verse the total number of ions $N$. The dotted curve
represents the estimated uncertainty in SQL corresponding to the product
state, and the dash-dotted curve is the Heisenberg limit (HL) with the NOON
state. We take $T=5$ ms and $\protect\omega_{0}\approx 2\protect\pi \times 3$
GHz at the magnetic field $B=0.1$ Tesla.}
\label{Fig.3}
\end{figure}

\section{Conclusions}

In summary, we have shown how to adiabatically carry out a MZ interferometer
based upon multiple trapped ions with artificial nonlinear interactions.
Influence from decoherence has also been discussed in our treatment. Since
our interferometry proposal involves large numbers of trapped ions and works
by global operations, we argue that the proposal would be useful for
high-precision quantum metrology with trapped ions toward the Heisenberg
limit~\cite{Wineland2}.

We should also emphasize that, although our discussion above has only
focused on the case of $\Delta >$ 0 and $\lambda >$ 0 for a trapped-ion MZ
interferometer, our model enables to study spin squeezing in an ion trap by
setting $\Delta <$ 0 for $\lambda <$ 0, which could be done by changing the
directions of the laser beams $k_{A}^{\prime }$\ and $k_{B}^{\prime }$ to
make $B^{x}<0$~\cite{spin squeezing}. Our model also enables to investigate
interaction blockade of spin flip with trapped ions for weak $B^{x}$~\cite%
{EPL}.

\section*{ACKNOWLEDGMENTS}

This work is supported by the NBRPC under Grants No. 2012CB821305 and
2012CB922102, the NNSFC under Grants No. 10974225, 11075223 and 11004226,
the NCETPC under Grant No. NCET-10-0850 and the Fundamental Research Funds
for Central Universities of China.

\appendix

\section{Generation of the effective Hamiltonian}

Below, we show how to generate the effective Hamiltonian (2) in Section II.

First, we discuss how to simulate the nonlinear term $\lambda J_{z}^{2}$ in
the Hamiltonian $H_{eff}$. As indicated in Fig. 1(a), irradiating the ions
to the longitudinal COM mode by applying two non-copropagating laser beams
with the wavevector difference $k_{z}$ along the $z$-axis, we have the
Hamiltonian below in the rotating frame regarding $H_{0}=(\nu -\Delta
)a^{\dag }a$ under the rotating-wave approximation,
\begin{equation}
H_{2}=e^{iH_{0}t}H_{1}e^{-iH_{0}t}=\Delta a^{\dag }a-\frac{1}{\sqrt{N}}
\sum\nolimits_{i=1}^{N}\Omega _{z}\eta _{z}(a^{\dag }+a)\sigma _{z}^{i}.
\label{A1}
\end{equation}
Then applying a canonical transformation $e^{-S}H_{2}e^{S}$ with $%
S=\sum\nolimits_{i=1}^{N} \frac{\Omega _{z}\eta _{z}}{\sqrt{N}\Delta}
(a^{\dag }-a)\sigma _{z}^{j}$~\cite{canonical}, we obtain
\begin{equation}
H_{3}=\Delta a^{\dag }a-\frac{2\Omega _{z}^{2}\eta _{z}^{2}}{N\Delta}
\sum\nolimits_{i,j}^{N}\sigma _{z}^{i}\sigma _{z}^{j}.  \label{A2}
\end{equation}

Cooling the COM motion of the ions to its ground states, we obtain an
effective spin-spin interaction in the Hamiltonian,
\begin{equation}
\tilde{H}_{3}=-\lambda J_{z}^{2},  \label{A3}
\end{equation}
with $\lambda =8\Omega _{z}^{2}\eta _{z}^{2}/(N\Delta )$ and $%
J_{z}=\sum\nolimits_{i=1}^{N}\sigma _{z}^{i}/2$.

The generation of the transverse-field term $-B^{x}J_{x}$ requires
additional lasers. As shown in Fig. 1(b), we set $\omega _{A}^{\prime
}-\omega _{B}^{\prime } = \omega _{0}$, implying a resonant Raman process
with respect to the two ground spin states of the ions. Such operations
yield the transverse-field term~\cite{canonical}
\begin{equation}
\tilde{H}_{2}=-B^{x}\sum\nolimits_{i=1}^{N}\sigma _{x}^{i}/2=-B^{x}J_{x},
\label{A4}
\end{equation}
with $B^{x}$ an effective Rabi frequency~\cite{canonical,Bx} and $J_{x}=\sum
\nolimits_{i=1}^{N}\sigma _{x}^{i}/2$.

If the Raman beams for generating the transverse-field term ($-B^{x}J_{x}$)
are non-resonant, i.e. the frequency difference of the two lasers $\omega
_{A}^{\prime \prime }-\omega _{B}^{\prime \prime }=\omega _{0}-\delta $ with
the detuning $\delta $, one can obtain the longitudinal-field term,
\begin{equation}
\tilde{H}_{1}=\delta \sum\nolimits_{i=1}^{N}\sigma _{z}^{i}/2=\delta J_{z},
\label{A5}
\end{equation}
see the schematic diagram Fig. 1(b). 

Since the manipulation on the transverse COM mode is independent of the
longitudinal COM mode, the total effective Hamiltonian reads as
\begin{equation}
H_{eff}=\tilde{H}_{1}+\tilde{H}_{2}+\tilde{H}_{3}=\delta
J_{z}-B^{x}J_{x}-\lambda J_{z}^{2}. \label{A6}
\end{equation}\\

\section{Calculation of $\Delta\protect\omega$}

For simplicity, we only consider the decoherence of the system during a free
evolution. Suppose the system to be prepared in the maximally path-entangled
state $\rho _{p}(0)=(\left\vert \frac{N}{2},\frac{N}{2}\right\rangle
+\left\vert \frac{N}{2}, -\frac{N}{2}\right\rangle )(\left\langle \frac{N}{2}%
,\frac{N}{2}\right\vert +\left\langle\frac{N}{2}, -\frac{N}{2}\right\vert
)/2 $, where $\left\vert \frac{N}{2},\frac{N}{2} \right\rangle $ and $%
\left\vert \frac{N}{2},-\frac{N}{2} \right\rangle$ stand for all ions in $%
\left\vert\uparrow \right\rangle$ and $\left\vert \downarrow \right\rangle$,
respectively. After a time duration of T, the system evolves into
\begin{eqnarray}
\rho _{p}(\omega _{0},T) &=&\frac{1}{2}\left(\left\vert \frac{N}{2}, \frac{N%
}{2}\right\rangle \left\langle \frac{N}{2},\frac{N}{2}\right\vert\right.
\notag \\
&&+\left\vert \frac{N}{2},-\frac{N}{2}\right\rangle \left\langle \frac{N}{2}%
,-\frac{N}{2}\right\vert  \notag \\
&&+e^{-\gamma TN^{2}+i\omega _{0}TN}\left\vert \frac{N}{2},-\frac{N}{2}%
\right\rangle \left\langle \frac{N}{2},\frac{N}{2}\right\vert  \notag \\
&&\left. +e^{-\gamma TN^{2}-i\omega _{0}TN}\left\vert \frac{N}{2},\frac{N}{2}%
\right\rangle \left\langle \frac{N}{2},-\frac{N}{2}\right\vert \right).
\notag \\
&&  \label{B1}
\end{eqnarray}
According to~\cite{estimation2}, the quantum Fisher information is given as
\begin{equation}
F_{Q}=Tr\left[\rho _{p}(\omega _{0},T)A^{2}\right]=N^{2}T^{2}e^{-2\gamma
TN^{2}},  \label{B2}
\end{equation}
where the Hermitian operator $A$ is the ``symmetric logarithmic derivative"
with its matrix elements defined as $A_{ij}=2[\rho_{p}^{\prime }(\omega
_{0},T)]_{ij}/(p_{i}+p_{j})$, $\rho _{p}^{\prime}(\omega _{0},T)=$ $\partial
\rho _{p}(\omega _{0},T)/\partial T$, with $p_{i}$ and $p_{j}$ the
eigenvalues of $\rho _{p}(\omega _{0},T)$, conditional on $p_{i} + p_{j}=0$
and $A_{ij}=0$.

The frequency uncertainty $\Delta \omega$ satisfies the quantum Cram\'{e}%
r-Rao bound, \cite{bound}
\begin{equation}
\Delta \omega \geq \frac{1}{\sqrt{kF_{Q}}}=\frac{1}{\sqrt{k}NTe^{-\gamma
TN^{2}}},  \label{B3}
\end{equation}
in the measurements on a set of $k$ probes.

\end{document}